\documentclass[twocolumn]{aastex62}

\usepackage{subfigure}
\usepackage{wrapfig}
\usepackage{gensymb}

\graphicspath{{./}{figures/}}

\received{November 15, 2019}
\revised{July 7, 2020}
\accepted{July 8, 2020}
\submitjournal{ApJ}

\shorttitle{Evryscope $\&$ \textit{K2} TRAPPIST-1 Superflare Occurrence}
\shortauthors{Glazier et al.}

\begin{document}
\title{Evryscope and \textit{K2} Constraints on TRAPPIST-1 Superflare Occurrence and Planetary Habitability}

\correspondingauthor{Amy Glazier}
\email{aglazier@unc.edu}

\author[0000-0001-9981-4909]{Amy L. Glazier}
\affil{Department of Physics and Astronomy, University of North Carolina at Chapel Hill, Chapel Hill, NC 27599-3255, USA}

\author[0000-0002-0583-0949]{Ward S. Howard}
\affil{Department of Physics and Astronomy, University of North Carolina at Chapel Hill, Chapel Hill, NC 27599-3255, USA}

\author[0000-0002-6339-6706]{Hank Corbett}
\affil{Department of Physics and Astronomy, University of North Carolina at Chapel Hill, Chapel Hill, NC 27599-3255, USA}

\author[0000-0001-9380-6457]{Nicholas M. Law}
\affil{Department of Physics and Astronomy, University of North Carolina at Chapel Hill, Chapel Hill, NC 27599-3255, USA}

\author[0000-0001-8791-7388]{Jeffrey K. Ratzloff}
\affil{Department of Physics and Astronomy, University of North Carolina at Chapel Hill, Chapel Hill, NC 27599-3255, USA}

\author[0000-0002-4227-9308]{Octavi Fors}
\affil{Department of Physics and Astronomy, University of North Carolina at Chapel Hill, Chapel Hill, NC 27599-3255, USA}
\affil{Institut de Ci\`encies del Cosmos (ICCUB), Universitat de Barcelona, IEEC-UB, Mart\'{\i} i Franqu\`es 1, E08028 Barcelona, Spain}

\author{Daniel del Ser}
\affil{Department of Physics and Astronomy, University of North Carolina at Chapel Hill, Chapel Hill, NC 27599-3255, USA}
\affil{Institut de Ci\`encies del Cosmos (ICCUB), Universitat de Barcelona, IEEC-UB, Mart\'{\i} i Franqu\`es 1, E08028 Barcelona, Spain}

\begin{abstract}
The nearby ultracool dwarf TRAPPIST-1 possesses several Earth-sized terrestrial planets, three of which have equilibrium temperatures that may support liquid surface water, making it a compelling target for exoplanet characterization. TRAPPIST-1 is an active star with frequent flaring, with implications for the habitability of its planets. Superflares (stellar flares whose energy exceeds $10^{33}$ erg) can completely destroy the atmospheres of a cool star's planets, allowing ultraviolet radiation and high-energy particles to bombard their surfaces. However, ultracool dwarfs emit little ultraviolet flux when quiescent, raising the possibility of frequent flares being necessary for prebiotic chemistry that requires ultraviolet light. We combine Evryscope and \textit{Kepler} observations to characterize the high-energy flare rate of TRAPPIST-1. The Evryscope is an array of 22 small telescopes imaging the entire Southern sky in \textit{g'} every two minutes. Evryscope observations, spanning 170 nights over 2 years, complement the 80-day continuous short-cadence K2 observations by sampling TRAPPIST-1's long-term flare activity. We update TRAPPIST-1's superflare rate, finding a cumulative rate of $4.2^{+1.9}_{-0.2}$ superflares per year. We calculate the flare rate necessary to deplete ozone in the habitable-zone planets' atmospheres, and find that TRAPPIST-1's flare rate is insufficient to deplete ozone if present on its planets. In addition, we calculate the flare rate needed to provide enough ultraviolet flux to power prebiotic chemistry. We find TRAPPIST-1's flare rate is likely insufficient to catalyze some of the Earthlike chemical pathways thought to lead to RNA synthesis, and flux due to flares in the biologically relevant UV-B band is orders of magnitude less for any TRAPPIST-1 planet than has been experienced by Earth at any time in its history.
\end{abstract}

\keywords{planets and satellites: terrestrial planets --- 
planetary systems --- stars: activity --- stars: flare --- stars: individual (TRAPPIST-1)}


\section{Introduction} \label{sec:intro}

TRAPPIST-1 (2MASS J23062928$\--$0502285) is an ultracool dwarf of spectral type M8V with seven terrestrial planets, three of which have equilibrium temperatures that may sustain surface liquid water \citep{Gillon2016, Gillon2017}. With Earth-sized planets in its habitable zone, where temperatures are not hot enough to cause water loss nor too cold for greenhouse warming to maintain temperatures conducive for liquid water \citep{Ravi2013}, the TRAPPIST-1 system is a compelling target for exoplanet characterization. However, a planet's location in the habitable zone of its star is not sufficient to guarantee that the planet will indeed have liquid water on its surface. Habitability depends on a multitude of characteristics of the planet and its host star, such as atmospheric composition of the planet and flare activity of the star \citep{Shields2016}. Like many ultracool dwarfs (e.g., \citealt{Schmidt2016}, \citealt{Paudel2018}, \citealt{Rodriguez2018}, \citealt{Schmidt2019}), TRAPPIST-1 flares on a regular basis, which may affect its planets' ability to harbor life. Large flares from the star have been observed during the \textit{K2} observing campaign \citep{Vida2017, Paudel2018}.

The UV radiation from a single stellar flare of an ultracool dwarf may not have a significant effect on planetary atmospheres, but the concomitant flux of high-energy particles can reduce the ozone column depth of an Earthlike planet's atmosphere by as much as 94$\%$ \citep{Segura2010}. While a planet's ozone column can recover from such a flare, the process takes years \citep{Loyd2018, Tilley2019, Ward2018}, leaving the planet's surface exposed to increased UV radiation that can be harmful for life. Extremely large flares can completely destroy ozone in an Earthlike planet's atmosphere, potentially allowing the planet's surface to be sterilized by UV radiation, although such events are rare in comparison to smaller flares \citep{Loyd2018}.

In its quiescent state, TRAPPIST-1 does not emit enough particle nor UV flux to threaten life on its planets; however, during extreme flare events, this is no longer the case, and incident radiation at planetary surfaces can reach lethal levels \citep{Yamashiki2019}. Furthermore, depending on the characteristics and alignment of TRAPPIST-1's magnetic field, its planets may be subjected to as much as six orders of magnitude greater proton flux than Earth experiences \citep{Fraschetti2019}. An analysis of the UV and X-ray flux experienced by the TRAPPIST-1 planets, in relation to their escape velocities, has shown that the activity of the star may have been sufficient to strip atmospheres and oceans alike \citep{Roettenbacher2017}.

Since flares can profoundly impact planetary habitability, we seek to constrain the flare rate of TRAPPIST-1 in order to quantify the effects of the star's flares on its planets. \textit{Kepler} observations of TRAPPIST-1 provided continuous coverage of the star over 80 days \citep{Vida2017, Paudel2018}. The Evryscope has provided continuous night-time all-sky monitoring since its deployment in 2015, and complements the short-term observations of \textit{Kepler} by characterizing TRAPPIST-1's long-term flare activity. Extremely large flares are within the Evryscope's ability to observe, even for faint sources normally undetectable by the system, allowing the Evryscope to characterize the long-term occurrence rate for those flares that are most damaging to potentially habitable planets. 

Long-term observations are particularly important, as some studies have found evidence for both long- and short-term stellar activity cycles on the order of a few years in stars similar to TRAPPIST-1 \citep{Baliunas1996, Olah2009, Mascareno2016, Brandenburg2017}. Although observations of TRAPPIST-1 have suggested it is unlikely to undergo solar-like activity cycles itself \citep{Dmitrienko2018, Roettenbacher2017}, observing stars over timescales of years can also increase the odds of observing higher-energy flares that occur less frequently and may not otherwise be observed in a single continuous observing window on the order of months.

In this paper, we combine Evryscope observations of TRAPPIST-1 with observations of TRAPPIST-1 from the \textit{K2} observing campaign of \textit{Kepler} in order to better characterize the high-energy flare rate of TRAPPIST-1 and its potential impact on planetary environments. Section \ref{sec:obsv} provides information on the two sources of data, the Evryscope in \ref{subsec:evr} and \textit{Kepler} in \ref{subsec:K2}. Section \ref{sec:methods} details the TRAPPIST-1 flare search process and results obtained with the Evryscope. Section \ref{sec:FFD} presents the cumulative flare frequency distribution of TRAPPIST-1 and its calculated superflare rate. Section \ref{sec:habitability} discusses the implications of our findings for the possible habitability of the TRAPPIST-1 planets, focusing on the UV flux due to flares impacting its planets. We summarize our work in Section \ref{sec:end}. 

\section{Observations} \label{sec:obsv}

\subsection{The Evryscope} \label{subsec:evr}
The Evryscope consists of 22 small telescopes housed in a single dome known as ``the mushroom,'' arranged for coverage of almost the entire Southern sky \citep{Law2015}. From its location at Cerro-Tololo Inter-American Observatory (CTIO), the Evryscope images its 8520 square degree field of view every two minutes. The system takes images in the \textit{g'} band with a typical dark-sky limiting magnitude of \textit{g'} = 16 and a resolution of 13" per pixel \citep{Jeff2019}. Each night, the Evryscope tracks the sky continuously for two hours as it takes images, then ratchets back to its original position and resumes tracking for another two hours, repeating throughout the night. This cycle yields an average of $\sim$6 hr of continuous monitoring across the visible sky. 

A custom pipeline analyzes the Evryscope data set at real-time rates \citep{Law2016}. With each two-minute exposure, each camera takes a 30-Mpx image, which is saved as a FITS file and calibrated using a custom wide-field solver. The background is modeled and subtracted, then forced-aperture photometry is extracted based on known source positions in a reference catalog. Light curves are then generated for approximately 15 million sources across the Southern sky by differential photometry in small regions of the sky, using carefully selected reference stars and a range of apertures. Residual systematics are removed using two iterations of the SysRem detrending algorithm \citep{Tamuz2005}.

The Evryscope data set contains 9,210 images centered on the position of TRAPPIST-1, spanning June 2016 to June 2018. Fewer images exist for TRAPPIST-1 than for most stars in the Evryscope data because of TRAPPIST-1's location in the sky: at a declination of $-5\degree$, TRAPPIST-1 is near the northern edge of the Evryscope's field of view, where there is less camera coverage than for other sources. From the Evryscope's location at CTIO, TRAPPIST-1 is observable from mid-June to late November. Given the observing constraints, images of TRAPPIST-1's position were taken at two-minute cadence for an average of two hours on each observable night. TRAPPIST-1 was observed for a total of 170 nights during the two years of Evryscope coverage, or 12.8 days continuous-equivalent over two years.

TRAPPIST-1 is not bright enough to be visible in Evryscope data when quiescent; at magnitude \textit{g'} = 19.3 \citep{Chambers2016}, TRAPPIST-1 is much dimmer than the typical Evryscope limiting magnitude of \textit{g'} $\approx$ 16. However, a sufficiently large flare would increase TRAPPIST-1's brightness to match or exceed the sky background in Evryscope data. Flares of 3-11 magnitudes have been observed from mid- to late M-dwarfs \citep{Paudel2018, Schmidt2016}. Previous flare searches with the Evryscope have detected a number of flares in this amplitude range \citep{Ward2019}. If originating from TRAPPIST-1, such flares would be visible in Evryscope images of the star; Section \ref{subsec:minflare} details how we determine constraints on the observability of bright flares from TRAPPIST-1. In order to search for flares from TRAPPIST-1 with Evryscope, we sequence all Evryscope images of TRAPPIST-1's location and analyze them as described in the following sections.

\subsection{Kepler} \label{subsec:K2}
\textit{Kepler} observed the TRAPPIST-1 system in its short-cadence mode for one 80-day cycle, spanning December 2016 to March 2017; while these observations cover the same time range in which the Evryscope observed TRAPPIST-1, all \textit{K2} observations took place during the time of year when TRAPPIST-1 was not observable from CTIO. The flare frequency distribution (FFD; the cumulative flare occurrence rate as a function of flare energy, see e.g. \citealt{Gershberg1972}, \citealt{Lacy1976}, or \citealt{Davenport2016} for details) of TRAPPIST-1 has been calculated from \textit{Kepler} data by \cite{Vida2017} and \cite{Paudel2018}. Both \cite{Vida2017} and \cite{Paudel2018} find the same flares in the \textit{K2} light curve of TRAPPIST-1, although their flare energy calculations differ, with energies calculated from raw data by \cite{Vida2017} being $\sim$10 times greater than those calculated later by \cite{Paudel2018} using data reduced from the \textit{K2} pipeline. We conduct our analysis with the more recent data of \cite{Paudel2018}.

\section{Methods} \label{sec:methods}

\subsection{Image sequence of TRAPPIST-1} \label{subsec:movie}

\begin{figure*}[t!]
    \centering
    \subfigure
    {
        \includegraphics[width=2.2in]{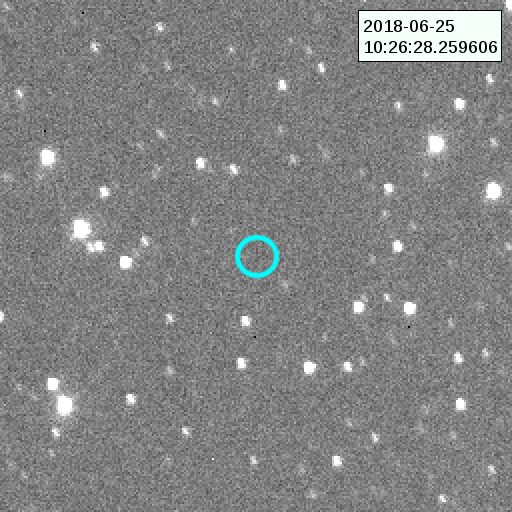}
    }
    \subfigure
    {
        \includegraphics[width=2.2in]{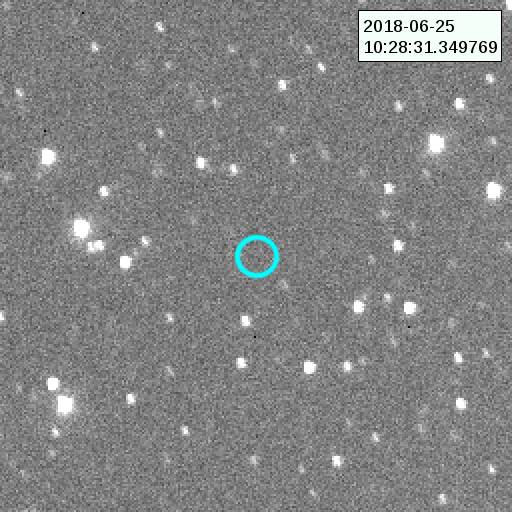}
    }
    \subfigure
    {
        \includegraphics[width=2.2in]{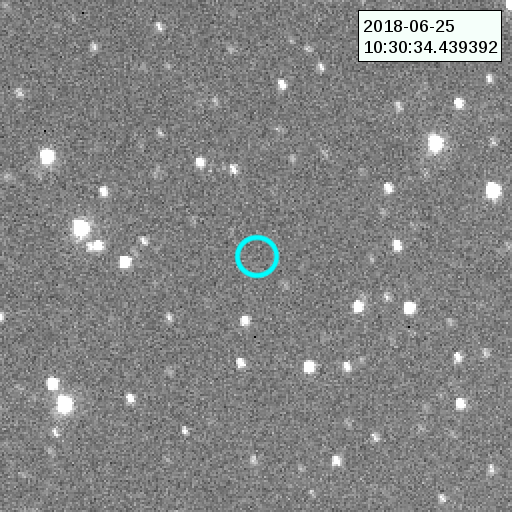}
    }
    \caption{An example set of frames from the image sequence of TRAPPIST-1. Typical PSFs for sources in the image span 2 to 4 pixels (26 to 52 arcsec). The expected position of TRAPPIST-1 is circled at center in each frame. The date and time (UT) of the image are given at upper right in each frame.}
    \label{fig:img_seq}
\end{figure*}

\begin{figure*}[t!]
    \centering
    \subfigure
    {
        \includegraphics[width=2.2in]{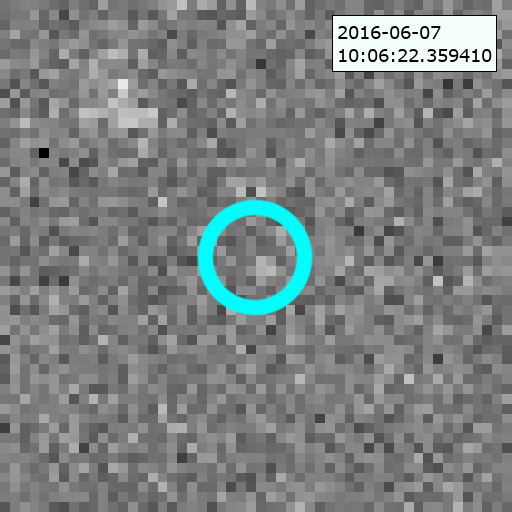}
    }
    \subfigure
    {
        \includegraphics[width=2.2in]{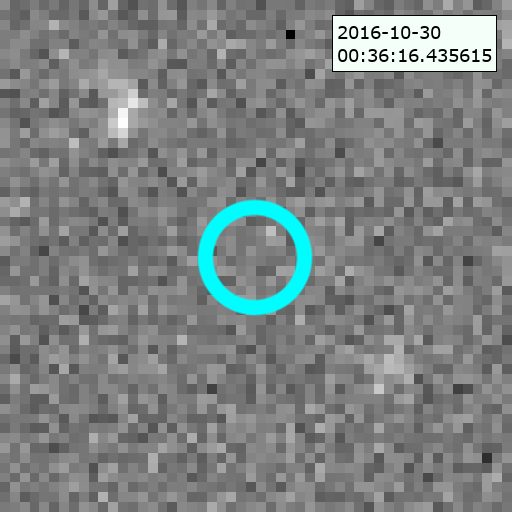}
    }
    \subfigure
    {
        \includegraphics[width=2.2in]{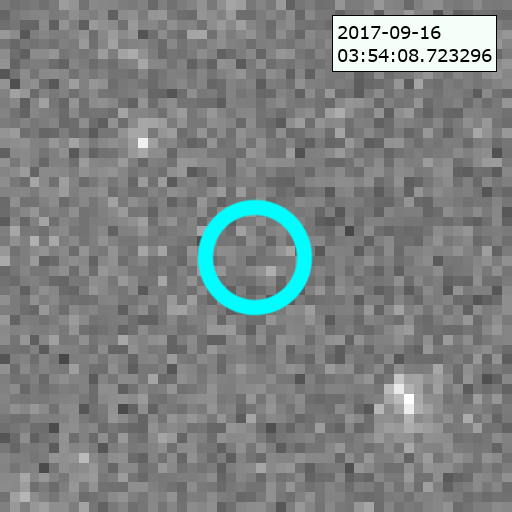}
    }
    \caption{Candidate flares observed in image sequence of TRAPPIST-1. The region of expected TRAPPIST-1 emission is circled at center in each. These images are zoomed in relative to the scale of those in Figure \ref{fig:img_seq}, showing the central five percent of the image in which each flare candidate occurs. Left: Candidate at 5 o'clock position in circle. Center: Candidate at 1 o'clock position in circle. Right: Candidate at 5 o'clock position in circle.}
    \label{fig:flare_cands}
\end{figure*}

Using a custom \verb|python| routine, we sequenced images of TRAPPIST-1 from the Evryscope for playback. By visual inspection of the image sequence, we could detect any excess emission indicative of flares at TRAPPIST-1's expected location in the image. We subtracted each image's median value from every pixel in the image to remove nightly variations in brightness due to changing sky background levels (caused by, for example, higher background from moonlight). We scaled the image contrast such that background noise -- and thus excursions above it -- was visible. Using the World Coordinate System (WCS) solutions provided by the Evryscope pipeline (precise at the arcsecond, or 0.1-pixel, level), we marked the position of expected TRAPPIST-1 emission in each image. We then converted all images from FITS to PNG and sequenced them at a frame rate of 2 frames/sec using \verb|FFmpeg|. An example set of consecutive frames from the image sequence is shown in Figure \ref{fig:img_seq}.

We performed an initial visual search of the image sequence to identify image regions with significant deviations from noise. Since each image is centered on TRAPPIST-1's right ascension and declination (after correcting for the star's proper motion), any emission from the star is expected to lie within a few pixels of the image center, in the region outlined in red in each image of Figure \ref{fig:img_seq}. Three flare candidates were visible in the image sequence; these candidates are shown in Figure \ref{fig:flare_cands}. The detection levels of the three candidates are $3.0\sigma$, $1.8\sigma$, and $2.0\sigma$, respectively, above the local noise floor. These candidates represent the most significant deviations from local noise found across the entire duration of the image sequence, but are not necessarily significant events as discussed below. Each candidate flare image was inspected by at least two people, and checked against the following criteria:
\begin{enumerate}
    \item Is the flare candidate more than one pixel in extent?
    \item Is the flare candidate's PSF shape consistent with nearby stars and signal level compared to the local image background?
    \item Is the flare candidate's brightness a 3$\sigma$ variation from the noise?
    \item Is the detected peak position of the flare candidate close enough to the expected location of TRAPPIST-1 to be likely to have originated from the star?
    \item Is the flare candidate visible in at least two consecutive images?
\end{enumerate}
We require the final criterion in order to ensure secure detections and to reduce false positives from fast transients, satellite trails, noise fluctuations, and other phenomena unrelated to TRAPPIST-1. In order to be detectable, the integrated flux of a flare must exceed the Evryscope's limiting magnitude for at least two minutes. Large flares detectable by the Evryscope likely persist for more than two minutes \citep{Ward2019}, and most have multiple peaks \citep{Davenport2014}, so a real flare would likely be visible in consecutive images.

The first flare candidate, shown at left in Figure \ref{fig:flare_cands}, meets all criteria except the last -- it is not visible in the frame immediately before its appearance, and there is no frame immediately following its appearance due to sunrise. The second candidate, at center in Figure \ref{fig:flare_cands}, also does not appear in the frame before or after its appearance, nor does its detection level ($1.8\sigma$) meet the third criterion and so it is rejected. The third candidate, at right in Figure \ref{fig:flare_cands}, does not appear in the previous frame and appears to move downward by 3 pixels in the next frame; however, its detection level ($2.0\sigma$) does not meet the third criterion, and its motion in the next frame is consistent with the local noise background, so it is likely caused by random noise fluctuations.

Assuming that a flare must be at least as bright as the Evryscope limiting magnitude \citep{Jeff2019} in order to be visible, we estimate the expected number of false-positive flares using Gaussian statistics. Given the number of images observed, we would expect a 3$\sigma$ deviation from the background far more often than the number of candidates actually observed. This suggests that our formal detection criterion is more stringent than 3$\sigma$, as expected given the extra filtering of candidates we applied. Detecting only three flare candidates is consistent with the false positive estimate of 2.3 $\pm$ 1.5 flares one obtains assuming our detection limits are 4$\sigma$ instead of 3$\sigma$. While we cannot exclude the possibility that each candidate was a real flare from the TRAPPIST-1 system, we cannot attach any significant probability of realness with only a single data point for all three candidates and a number of occurrences well within the expected number of false positive signals. No flares from TRAPPIST-1 were confirmed in the Evryscope data set.

\subsection{Evryscope Minimum Detectable Flare Energy} \label{subsec:minflare}
In order to determine the completeness limit of our method, we perform flare rejection and recovery simulations to find flare amplitudes and energies with 100$\%$ completeness. We can therefore establish a limit on how often flares of detectable energies can occur, given that we observed none. In order to do so, we run a series of Monte Carlo simulations to estimate the Evryscope minimum detectable flare energy for TRAPPIST-1.

\begin{figure}
    \centering
    \includegraphics[scale=0.4]{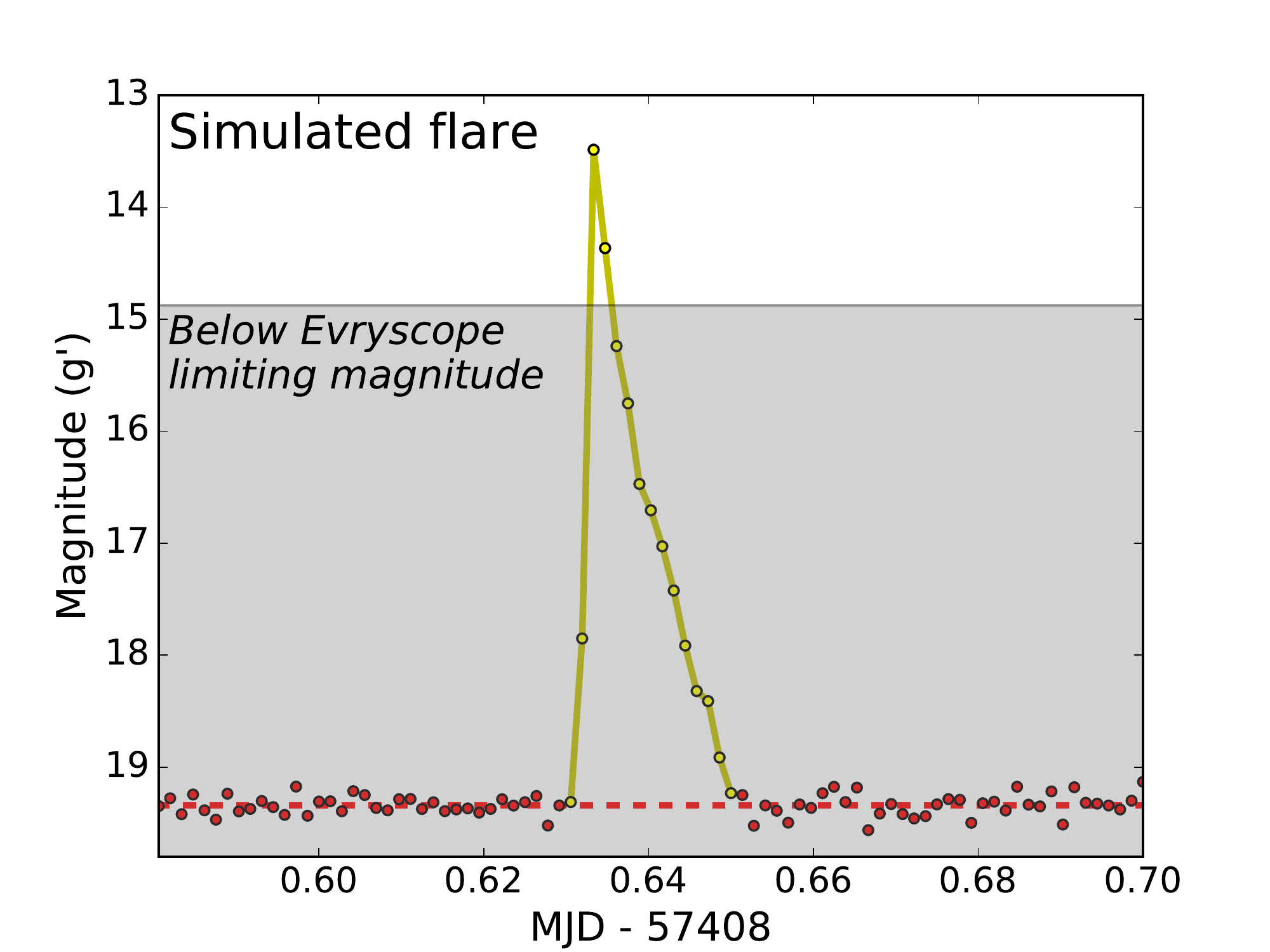}
    \caption{One of many simulated flares generated to determine the Evryscope minimum recoverable flare energy for TRAPPIST-1. A flare of this magnitude originating from TRAPPIST-1 would have been visible in Evryscope images of the star.}
    \label{fig:sim_flare}
\end{figure}

Since TRAPPIST-1 is not visible to Evryscope when quiescent, we simulate its light curve as a one-day-long array of magnitude data matching the Evryscope's 2-minute cadence, with values equal to TRAPPIST-1's quiescent \textit{g'} magnitude of 19.3. Following the methods described in \cite{Ward2018}, we inject simulated flares into the light curve to determine the Evryscope's flare recovery rate for TRAPPIST-1. For 10,000 trials, flares are generated using the template described in \cite{Davenport2014} with possible peak contrasts (i.e. difference in magnitude between flare peak and quiescence) of 6.0, 5.5, 5.0, 4.75, 4.5, 4.25, 4.0, 3.75, 3.5, 3.25, and 3.0 in \textit{g'}. An example simulated flare is shown in Figure \ref{fig:sim_flare}. 

\begin{figure}
    \centering
    \includegraphics[scale=0.4]{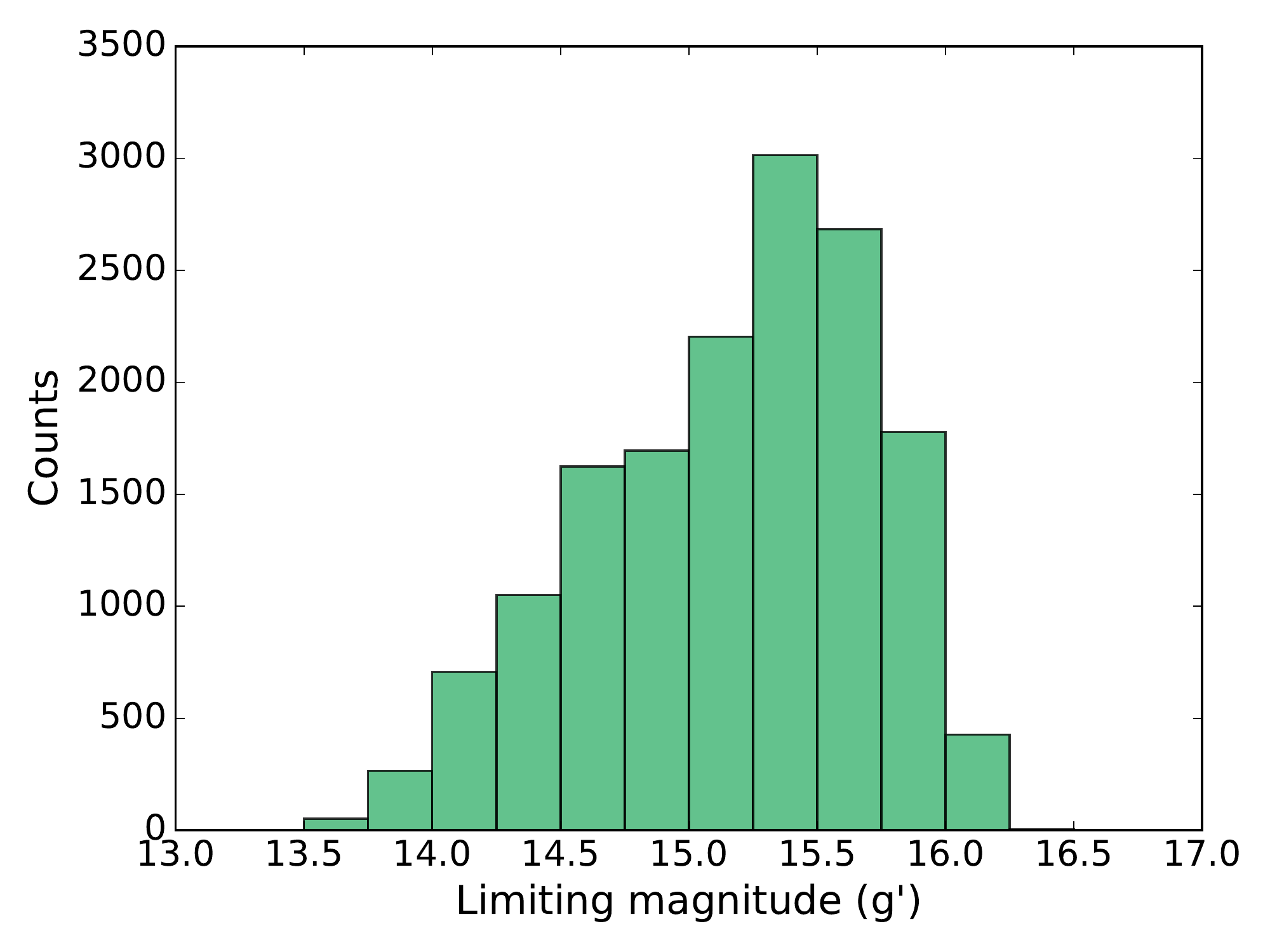}
    \caption{Distribution of Evryscope's 3-$\sigma$ limiting magnitudes measured at TRAPPIST-1's location over the past 3 years of Evryscope observations, including all sky, moon and cloud conditions.}
    \label{fig:lim_mag_hist}
\end{figure}

Flares are injected into the simulated light curve at randomly chosen epochs, and passed through checks to determine if at least two points of the flare are brighter than or equal to the Evryscope's limiting magnitude (to satisfy the final detection criterion given in Section \ref{subsec:movie}). The limiting magnitude in each trial is drawn randomly from a distribution of Evryscope's measured limiting magnitude values at TRAPPIST-1's position between June 2016 and June 2018, matching the time span covered by the Evryscope images analyzed in Section \ref{subsec:movie}. The Evryscope limiting magnitudes are calculated from a comparison of the signal-to-noise ratio of detection of hundreds of nearby stars to catalog magnitudes from APASS \citep{apass}; the full algorithm is described in \cite{Jeff2019}. Figure \ref{fig:lim_mag_hist} shows the distribution of measured limiting magnitudes. Simulated flares brighter than or equal to the Evryscope's limiting magnitude for two or more consecutive epochs are counted as successful recoveries; such flares would have been visible in images of TRAPPIST-1 had they occurred during Evryscope observations of the star. We define flare contrast, $\Delta\textit{g'}$, as the difference in \textit{g'} magnitude between the flare peak and the star's quiescent \textit{g'} magnitude.

\begin{figure}
    \centering
    \includegraphics[scale=0.5]{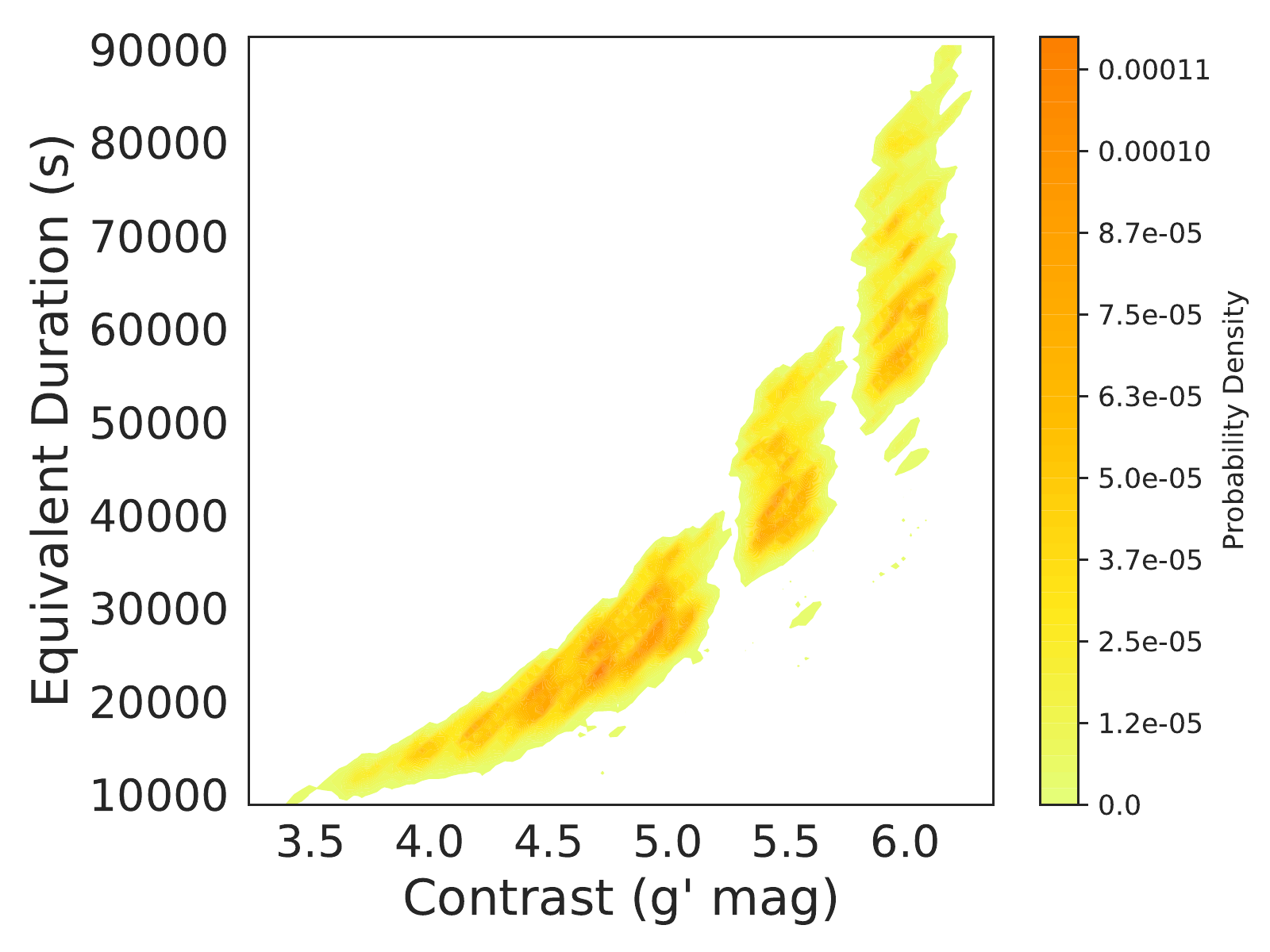}
    \caption{Equivalent durations (EDs) of recovered simulated flares as a function of flare contrast in \textit{g'}. The distribution's banana-like shape is due to the fact that a flare of a given ED could be a short-duration flare with high peak contrast, or a long-duration flare with a low peak contrast, such that both situations would have the same ED (area under the curve) but different peak contrasts.}
    \label{fig:bananas}
\end{figure}

\begin{figure}
    \centering
    \includegraphics[scale=0.5]{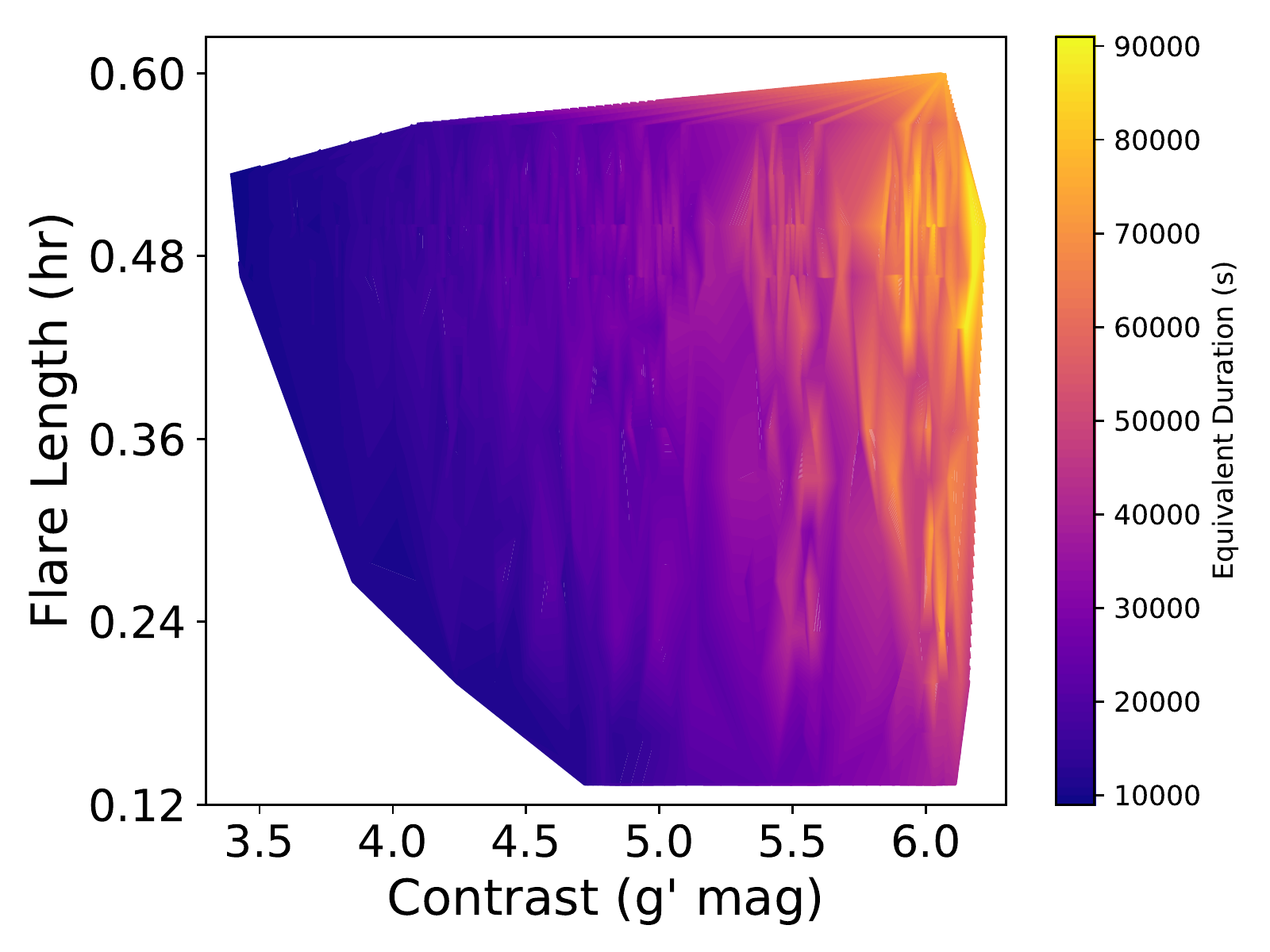}
    \caption{Length in hours of recovered simulated flares as a function of flare contrast in \textit{g'}, with flare equivalent duration (ED) denoted by color. The yellow region, concentrated primarily along the right-hand side of the plot, denotes higher EDs (thus higher overall flare energy), while the deep violet region along the left half of the plot denotes lower EDs (lower energy). High-amplitude and long-duration flares demonstrate the highest ED values.}
    \label{fig:length_vs_contrast}
\end{figure}

\begin{figure}
    \centering
    \includegraphics[scale=0.5]{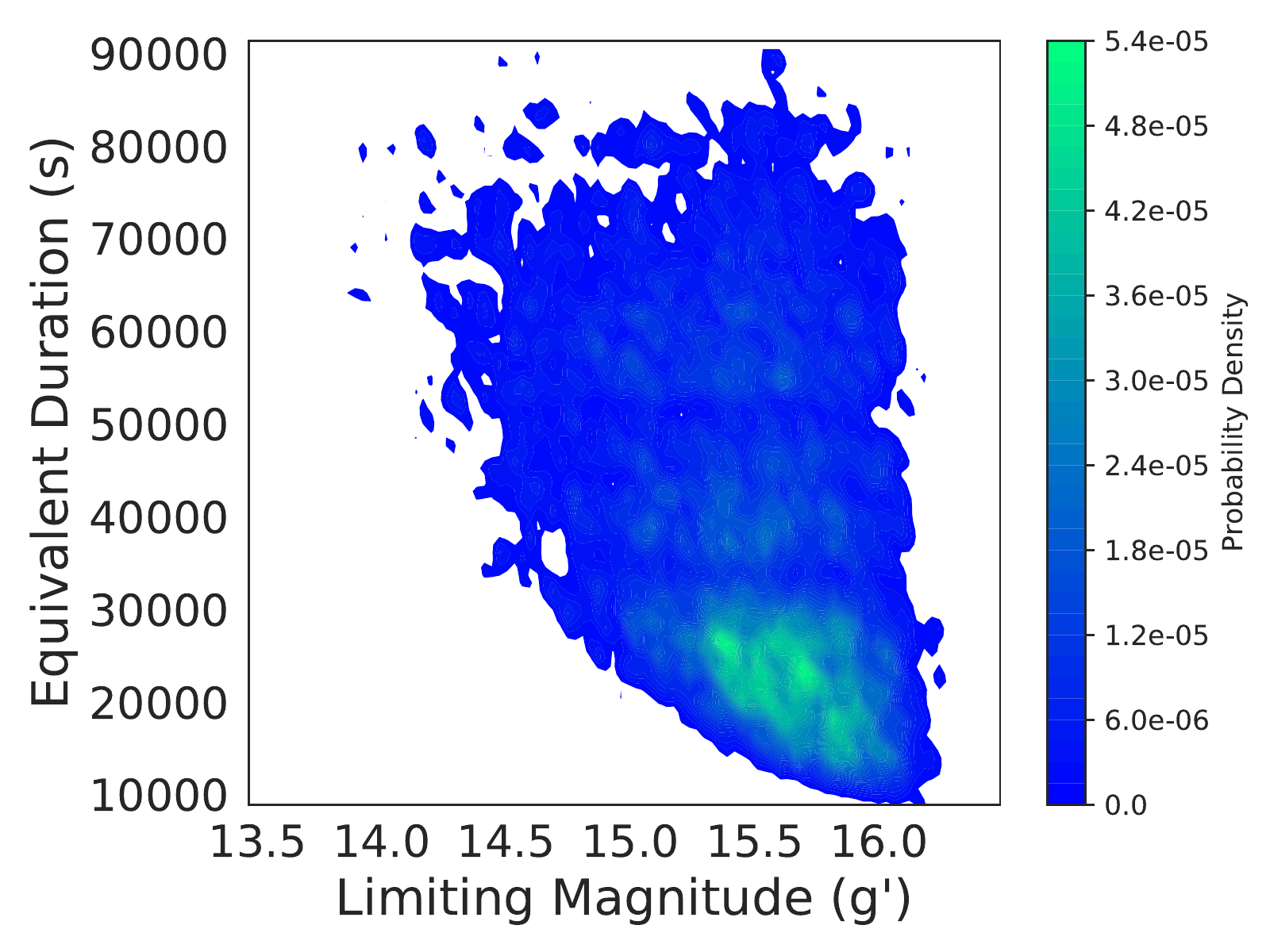}
    \caption{Equivalent durations (EDs) of recovered simulated flares as a function of limiting magnitude. Lower-energy flares can be detected if the limiting magnitude is dimmer; higher-energy flares are detectable over a wider range of limiting magnitudes, but occur less frequently than lower-energy flares. Most recovered flares are concentrated in the lower right (green) corner of the distribution, since that region overlaps the most common flare EDs and the most common limiting magnitude.}
    \label{fig:swoosh}
\end{figure}

The total energy radiated by a flare depends on both its energy output at its peak and how long the flare lasts; flares with lower peak energies but long durations can have the same total energy output as flares with higher peak energies but much shorter durations. The equivalent duration (ED) is a measure of total flare energy output that takes account of this degeneracy between peak energy and duration, by calculating the normalized flux of the flare integrated over the length of the flare, yielding the equivalent duration of the flare in units of seconds. Here, the ED is calculated for each simulated flare as a first step toward calculating the flare energy. We calculate the ED following the method of \cite{Hawley2014}. We use the trapezoidal rule to calculate ED as the ``area under the curve'' of each simulated flare, where the upper and lower limits of integration are the stop and start times of the flare, respectively. Using Figure \ref{fig:sim_flare} as an example, calculating the ED amounts to calculating the area enclosed by the quiescent emission (red points) as the x-axis and a curve tracing the flare emission (yellow points) as the enclosing function, from the flare start time to the flare end time. Figure \ref{fig:bananas} shows the ED distribution of simulated flares with sufficient contrast to be potentially visible in Evryscope data; the distribution's dropoff below $\Delta\textit{g'} \approx 3.5$ signifies that flares with contrasts less than $\Delta\textit{g'} \lesssim 3.5$ cannot boost TRAPPIST-1's quiescent magnitude of \textit{g'} = 19.3 brighter than the Evryscope limiting magnitude of \textit{g'} $\approx$ 16. Figure \ref{fig:length_vs_contrast} shows the distribution of recovered flare time durations as a function of contrast, and Figure \ref{fig:swoosh} shows the distribution of recovered flare EDs as a function of limiting magnitude.

We estimate the Evryscope minimum recoverable flare energy using the ED and limiting magnitude distributions as follows. 
\begin{enumerate}
    \item We take the median of the limiting magnitude distribution in Figure \ref{fig:lim_mag_hist}, finding that the Evryscope's median limiting magnitude, over all sky, moon and cloud conditions at TRAPPIST-1's position is \textit{g'} = 15.52.
    \item We check the simulated flare distribution against the median limiting magnitude obtained in step 1, to find the lowest flare contrast associated with that limiting magnitude. We take this contrast to be the average minimum recoverable flare contrast, since it occurs for the average value of limiting magnitude. The minimum flare contrast thus obtained is $\Delta\textit{g'} = 3.89$.
    \item We recover the minimum ED from the distribution of flare EDs as a function of contrast (Figure \ref{fig:bananas}, finding a minimum ED of 13000 s for the minimum contrast of $\Delta\textit{g'} = 3.89$ from step 2. 
    \item Using TRAPPIST-1's quiescent \textit{g'} magnitude (19.3) and its distance from Earth (12.1 pc), we calculate TRAPPIST-1's quiescent \textit{g'} luminosity ($10^{27.4}$ erg/s) using the distance-magnitude relation. We then multiply the quiescent \textit{g'} luminosity by the minimum recoverable ED to obtain the minimum recoverable \textit{g'} flare energy, $E_{min,\textit{g'}} = 10^{31.5}$ erg.
    \item Finally, we convert the minimum recoverable flare energy from the Evryscope \textit{g'} bandpass to bolometric energy using the energy partitions of \cite{OstenWolk2015}. To do so, we estimate the bolometric energy of a flare as being that of a 9000 K blackbody, finding that such a flare emits a fraction $f_{g'} = 0.19$ of its bolometric energy in the \textit{g'} bandpass. Division of the minimum recoverable flare energy in \textit{g'} by $f_{g'}$ yields the minimum recoverable bolometric flare energy. We find the Evryscope minimum detectable bolometric flare energy is $E_{min} = 10^{32.2}$ erg. We note that the canonical 9000 K blackbody temperature provides a lower limit for flares' energies; a higher-temperature flare blackbody, as has been measured for some larger flares (e.g. \citealt{Kowalski2010}) would result in more energy radiated at short wavelengths.
\end{enumerate}

\section{Cumulative Flare Occurrence Rates} \label{sec:FFD}

We refine previous estimates of TRAPPIST-1's flare occurrence rate by combining our nondetection of higher-energy flares with \textit{Kepler} observations of flares during the \textit{K2} observing campaign. Since no flares exceeding the Evryscope minimum recoverable flare energy were observed during the 12.8 continuous-equivalent days of Evryscope observations, we determine the higher-energy flare rate for TRAPPIST-1 must be lower than calculated using \textit{K2} data alone. This decrease may be caused by statistical effects or a true change in flare activity. We compute the FFD parameters similarly to the methods described in \cite{Davenport2016} and \cite{Ward2019}. We model the cumulative flare occurrence rate as a power law of the form $\log{\nu} = \alpha\log{E} + \beta$, where $\nu$ is the number of flares per day whose bolometric energies are greater than or equal to $E$ erg, $\alpha$ is the frequency at which flares occur, and $\beta$ is the y-intercept, or the number of flares per day whose energy is $E = 10^{0} = 1$ erg; the value of $\beta$ sets the overall flare rate. We fit this power law to the \textit{K2} flares observed in \cite{Paudel2018}, after converting \textit{Kepler} bandpass flare energies given therein to bolometric flare energies as described in Section \ref{subsec:minflare}, using the fraction of bolometric energy in the \textit{Kepler} bandpass ($f_{K2} = 0.16$; \cite{OstenWolk2015}) rather than $f_{g'}$. For \textit{K2} flares of energy $E < E_{min}$, we calculate the cumulative flare occurrence rate using the 80-day observation duration of \textit{K2}. For \textit{K2} flares of energy $E \geq E_{min}$, we calculate the cumulative flare occurrence rate using the total observation time (approximately 92.8 days) of both \textit{K2} (80 days) and Evryscope ($\sim$12.8 days continuous-equivalent over two years). 

\begin{figure}[t]
    \centering
    \includegraphics[scale=0.33]{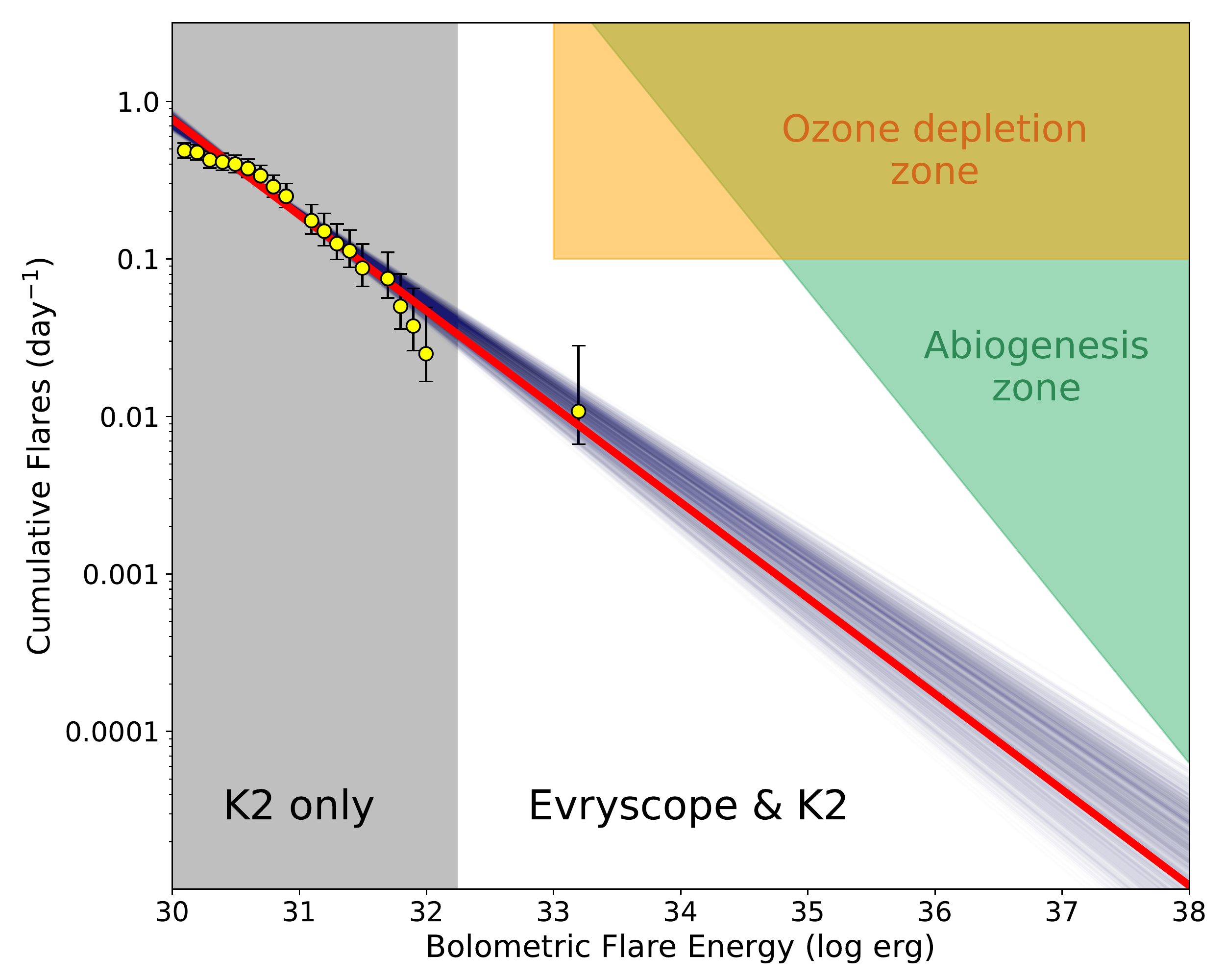}
    \caption{Updated flare frequency distribution of TRAPPIST-1. Yellow points are flares observed by \textit{K2}. No flares were observed by the Evryscope.}
    \label{fig:uh_oh}
\end{figure}

Figure \ref{fig:uh_oh} shows the new FFD for TRAPPIST-1, recomputed for flares exceeding the minimum recoverable Evryscope flare energy. We note this minimum energy is an approximation good to within an order of magnitude, sufficient for determining the adjusted superflare rate. The best-fit line to observed data, assuming a power law, is shown in red, with uncertainty in the fit illustrated by transparent dark blue lines. Ten thousand posterior draws from the flare distribution were used to determine the uncertainty in flare rates and best-fit parameters; we plot 2000 dark blue lines instead of all 10,000 for visual clarity. In order to calculate the uncertainty, we compute the frequency error bars as the $1\sigma$ confidence interval for \textit{K} flares observed in \textit{N} days. The frequency values are then randomly shuffled within their error bars to obtain the posterior draws and superflare rate. We compute the best-fit power law slope $\alpha$ and intercept $\beta$ as the mean of the distribution for each parameter, with uncertainties given by the $68\%$ and $95\%$ confidence intervals for each distribution. We find $\alpha = -0.61^{+0.01}_{-0.07}$ ($68\%$ confidence) and $-0.61^{+0.04}_{-0.09}$ ($95\%$ confidence), while $\beta = 18.1^{+2.2}_{-0.1}$ ($68\%$ confidence) and $18.1^{+2.8}_{-1.3}$ ($95\%$ confidence). Over the two years of Evryscope observation, we can exclude the possibility that the K2 detections were probing a particularly quiet time in TRAPPIST-1's activity cycle, but cannot reject that they were during a particularly active time. Had TRAPPIST-1 flared significantly more strongly than was observed by \textit{K2}, the Evryscope would have observed such a change.

The orange ozone depletion zone in Figure \ref{fig:uh_oh}, adapted from the analysis of \cite{Tilley2019}, marks the parameter space of cumulative flare rates and energies that lead to ozone loss for habitable-zone planets orbiting an M dwarf. The green abiogenesis zone, adapted from the analysis of \cite{Rimmer2018} and \cite{Gunther2019}, marks the region of flare rates and energies that result in enough UV flux to sustain prebiotic chemistry. We discuss each of these regions further in Section \ref{sec:habitability}.

We calculate the number of detectable flares predicted by Evryscope data alone, using the number of flares detected and the observing time. The observing time is calculated as the number of images of TRAPPIST-1 times the two-minute exposure time for each image. The number of flares detected is zero for that time span, with an upper limit given by a 1$\sigma$ binomial confidence interval. For flares of sufficiently high energy for the Evryscope to detect, this upper limit is 28 flares per year.

In conjunction with the \textit{K2} flare detections of \cite{Paudel2018}, the absence of detected Evryscope flares serves to downweight the overall cumulative occurrence rate for flares of energy $E \geq E_{min}$ compared to the rate calculated using \textit{K2} data alone. With only \textit{K2} data, at $68\%$ confidence, the resultant FFD would predict an annual cumulative occurrence rate of $\nu_{\log{E}\geq32.2} = 12.5^{+3.5}_{-0.3}$ flares per year for flares meeting or exceeding the Evryscope minimum recoverable energy of $\log{E_{min}} = 32.2$ log erg, and an annual cumulative superflare rate of $\nu_{\log{E}\geq33.0} = 4.4^{+2.0}_{-0.2}$ superflares per year. The new FFD in this work, derived from both \textit{K2} and Evryscope, yields an annual cumulative occurrence rate of $\nu_{\log{E}\geq32.2} = 12.1^{+3.5}_{-0.3}$ flares per year ($68\%$ confidence) or $12.1^{+4.9}_{-1.8}$ ($95\%$ confidence) that meet or exceed the Evryscope minimum recoverable flare energy, and an annual cumulative superflare occurrence rate of $\nu_{\log{E}\geq33.0} = 4.2^{+1.9}_{-0.2}$ superflares per year ($68\%$ confidence) or $4.2^{+2.8}_{-0.9}$ ($95\%$ confidence).

\section{Implications for Planetary Habitability} \label{sec:habitability}

In this section, we discuss the impact of TRAPPIST-1's flare rate on the potential habitability of its planets. We focus our analysis on flare-driven ozone depletion and abiogenesis, based on the cumulative UV flux from TRAPPIST-1's flares; our analysis is similar to that conducted by \cite{Gunther2019} in their study of a sample of TESS stars. Figure \ref{fig:uh_oh} shows the FFD of TRAPPIST-1, annotated with biologically relevant regions of the flare parameter space.

\subsection{Ozone Depletion} \label{subsec:oz_depl}
We investigate the effects of TRAPPIST-1's flares on ozone depletion by applying the findings of \cite{Tilley2019} to the TRAPPIST-1 system. In that work, \cite{Tilley2019} estimate the flare rates and energies for a general M dwarf that would deplete $\geq 99.99\%$ of the ozone column in a habitable-zone planet's atmosphere, given the probability of any given flare actually impacting a habitable-zone planet due to the geometry of the system. \cite{Tilley2019} simulate M-dwarf flares at a variety of energies and frequencies to study their effect on the ozone column of planets orbiting the M dwarf. They find that flares with energy $E \geq 10^{34}$ erg are the primary contributor to ozone depletion; assuming the probability of any flare impacting the planet is 8.3$\%$, they estimate that a flare rate of 0.4 such flares per day is sufficient to deplete $\geq 99.99\%$ of the ozone column, or a worse case of 0.1 such flares per day if the probability of flare impact is 25$\%$ \citep{Tilley2019}. 

Since our interest lies in determining whether the TRAPPIST-1 planets are in any danger of atmospheric ozone depletion from flares, we consider the case of 0.1 flares-per-day of $\geq 10^{34}$ erg as a lowest effective energy for planetary ozone depletion, and calculate whether the FFD of TRAPPIST-1 is consistent with this case. In Figure \ref{fig:uh_oh}, we plot the energy regime corresponding to ozone depletion in orange. 

\cite{Tilley2019}'s model assumes a habitable-zone planet at a distance of 0.16 au from its host star, whereas the TRAPPIST-1 habitable-zone planets are approximately an order of magnitude closer than that to their star. A distance decrease of one order of magnitude corresponds to a flux increase of two orders of magnitude; however, UV flux accounts for $\sim10\%$ of combined ozone loss, with particle bombardment accounting for the remainder \citep{Tilley2019}. Shifting the ozone depletion region in Figure \ref{fig:uh_oh} as much as two orders of magnitude lower in energy still does not intersect the FFD. TRAPPIST-1 has at least an order of magnitude too low a flare rate to cause $\geq 99.99\%$ atmospheric ozone loss for habitable-zone planets, even in the worst case scenario that assumes a 25$\%$ probability of any flare impacting a planet in the habitable zone. We conclude that TRAPPIST-1's current rate of flaring is unlikely to cause complete atmospheric ozone loss, if ozone is present in the atmospheres of its habitable-zone planets. 

\subsection{Prebiotic Chemistry} \label{subsec:chemmy_bits}
We investigate TRAPPIST-1's flares as catalysts for prebiotic chemistry, or abiogenesis, by applying the work of \cite{Rimmer2018} to the TRAPPIST-1 system. \cite{Rimmer2018} combine experimental lab chemistry with stellar physics to estimate how much energy due to flares would be necessary to power abiogenesis on Earthlike planets orbiting an M dwarf. \cite{Rimmer2018} determine reaction rates for hydroperoxide and cyanic acid, two reaction products necessary for the reaction pathway leading to synthesis of RNA pyrimidine nucleotides, the synthesis of which requires the input of 254-nm UV light in order to proceed. 

Noting that a sustainable prebiotic chemistry for supporting life requires at least a 50$\%$ yield of each product in the nucleotide synthesis pathway, \cite{Rimmer2018} derive a formula for the minimum cumulative rate $\nu$ of flares at a given U-band energy $E_{u}$ that provides the requisite yield. We use \cite{Rimmer2018}'s formula, expressed in cumulative flares per second for U-band flux,
\begin{equation}
    \nu(E_u) = {8.0 \times 10^{27}} \times E^{-1}_u
\end{equation}
and convert to erg/day in bolometric flux using blackbody conversion factors in \cite{OstenWolk2015}, obtaining an equation for the cumulative flare rate per day as a function of bolometric energy in erg,
\begin{equation}
    \nu(E_{bol}) = {6.6 \times 10^{33}} \times E^{-1}_{bol}
\end{equation}
for which abiogenesis can proceed at rates sufficient for life. We then determine whether the FFD of TRAPPIST-1 overlaps with this energy regime; Figure \ref{fig:uh_oh} shows the FFD does not intersect the region that would meet \cite{Rimmer2018}'s criteria for supporting abiogenesis. While TRAPPIST-1's current flare rate is unlikely to deplete enough ozone from its planets to endanger any extant surface life, its current flare rate is also unlikely to provide enough UV flux to sustain prebiotic chemistry necessary for the synthesis of RNA nucleotides.

\subsection{UV-B Flux at TRAPPIST-1 Planets versus Earth} \label{subsec:uvb}
We estimate the cumulative annual UV flux from flares incident on each planet in the TRAPPIST-1 system, and compare to the quiescent UV flux Earth has experienced from our Sun. We do this in order to contextualize the flare UV flux, to analyze how it compares to the UV flux received by the only planet currently known to be inhabited. We focus on the UV-B (280-315 nm) region of the spectrum. Since UV-B light is energetic enough to damage biological tissues while being absorbed much less by Earthlike atmospheres than the higher-energy UV-C \citep{Rugheimer2015}, the UV-B flux reaching TRAPPIST-1's planets is particularly relevant for considerations of its surface habitability. 

\begin{figure*}[t]
    \centering
    \includegraphics[scale=0.44]{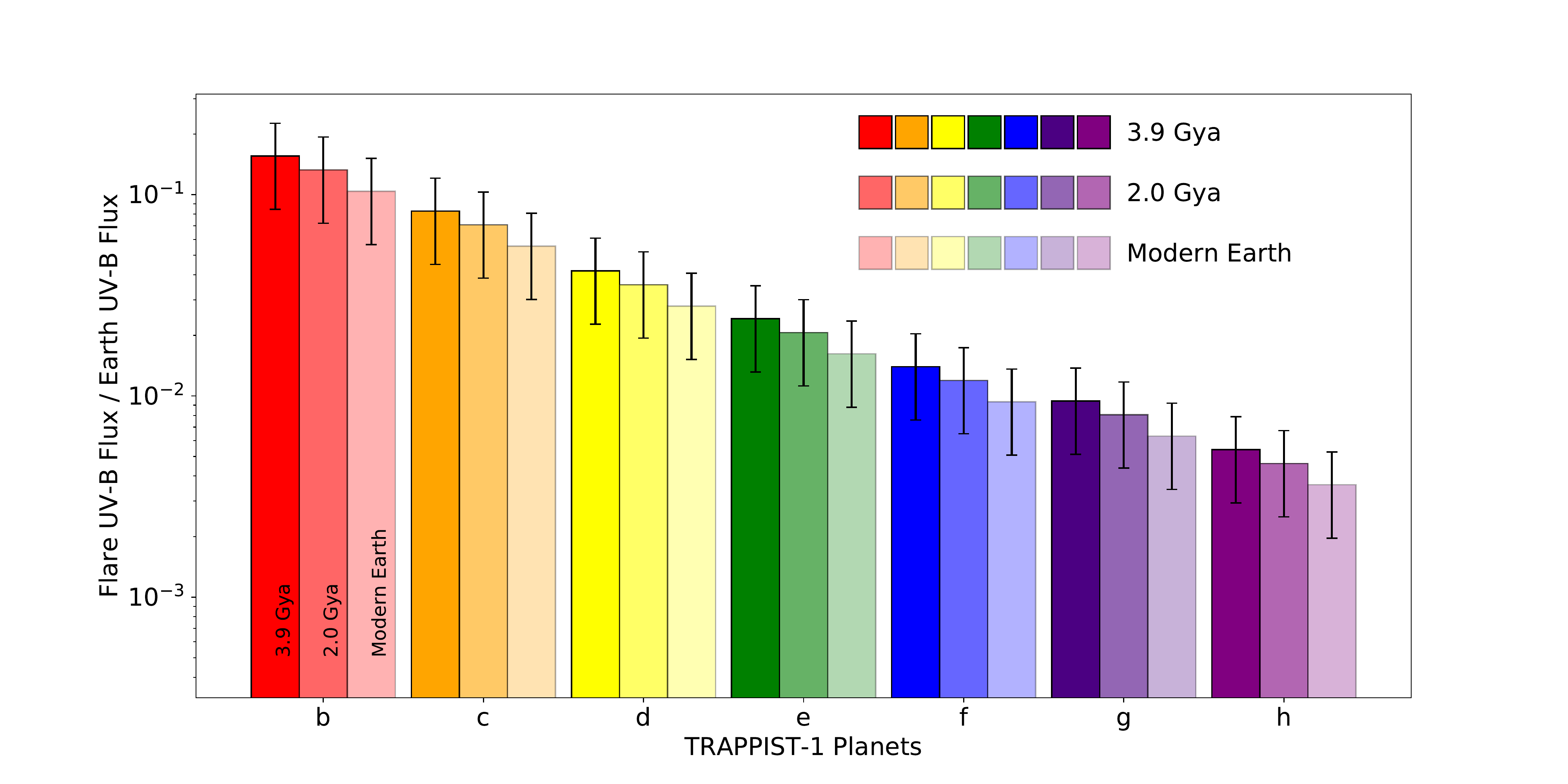}
    \caption{Cumulative annual top-of-atmosphere (TOA) flare UV-B flux at each of TRAPPIST-1's planets, as a fraction of Earth's cumulative annual TOA UV-B flux from our Sun for each of three epochs in Earth's history. The height of each bar gives the cumulative annual TOA flare UV-B flux for each TRAPPIST-1 planet as a fraction of that of Earth. The highest-saturation colors correspond to TRAPPIST-1 planets' UV-B flux ratio 3.9 Gya; the medium-saturation colors correspond to the proterozoic era 2.0 billion years ago; and the lowest-saturation colors correspond to modern Earth. Flux received decreases from left to right because the planets are ordered by increasing distance from the host star, with the closest planet (b) first in red and the farthest (h) last in violet. The trend is roughly linear because the planets are in a resonant chain.}
    \label{fig:rainbows}
\end{figure*}

We sample the FFD in Figure \ref{fig:uh_oh} within the uncertainty range of the power-law fit, generating a distribution of the number of flares per day meeting or exceeding some minimum energy. We then generate a distribution of cumulative annual flare energies by the following process:
\begin{enumerate}
    \item Select one number of flares per day at random from the distribution of number of flares per day, convert from flares per day to flares per year, and round to the nearest integer. For example: 0.042 flares per day $\rightarrow$ 15.33 flares per year $\approx$ 15 flares per year.
    \item Sample that number of flares from the FFD by generating flares of random energies, accepting those whose energies are consistent with values expected for their corresponding number of flares per day from the FFD, and rejecting those that are not consistent. We thereby obtain a distribution of possible energies for that number of flares per year. Example: Sample 15 flares from the FFD, yielding a list of 15 bolometric flare energies consistent with the FFD.
    \item Sum up the flare energies from step 2, yielding a single cumulative flare energy per year -- or cumulative annual flare energy.
    \item Repeat steps 1-3 for a total of 10,000 trials, yielding a distribution of cumulative annual bolometric flare energies.
\end{enumerate}
We take the median of the distribution to be the estimated cumulative annual bolometric flare energy, and assign uncertainties by taking the 1$\sigma$ confidence interval of the distribution.

To estimate the cumulative annual UV-B flare flux, we represent the spectral energy distribution of a flare as a 9000-K blackbody, which models the white-light continuum emission in flares while providing empirical fits for flare emission lines \citep{Hawley2003, Kowalski2010}. Modeling flares as blackbodies allows us to estimate their energy output in the UV-B region of the spectrum, using blackbody-dependent conversion relations for the fraction of bolometric energy encompassed by a given spectral band. Using this model, we convert the cumulative annual bolometric flare energies to cumulative annual UV-B flare energies.

From the cumulative annual UV-B flare energies, we estimate the cumulative annual UV-B flux received by each planet in the TRAPPIST-1 system, using their distances \citep{Grimm2018} to calculate the flare UV-B flux incident at the top of each planet's atmosphere. In doing so, we assume all flares hit each planet, making our estimate an upper limit on the cumulative annual UV-B flux received by each planet due to flares from TRAPPIST-1. Lastly, to contextualize our results, we convert the cumulative annual top-of-atmosphere (TOA) UV-B flux at each planet to Earth-relative TOA UV-B flux, dividing each TRAPPIST-1 planet's TOA UV-B flux by the values of TOA UV-B flux for Earth given in \cite{Rugheimer2015}. Using the tables provided by \cite{Rugheimer2015}, we compare our estimates of flare TOA UV-B flux to the quiescent TOA UV-B flux received by Earth during three epochs: early Earth 3.9 billion years ago before the rise of life, the proterozoic era 2.0 billion years ago as oxygen levels rose in Earth's atmosphere, and modern Earth. 

Figure \ref{fig:rainbows} summarizes the estimated Earth-relative UV-B fluxes at each TRAPPIST-1 planet. As a whole, the TOA UV-B flare flux for all TRAPPIST-1 planets pales in comparison to that which Earth has experienced in any epoch. TRAPPIST-1b, the closest planet to the star, receives an order of magnitude less UV-B flux from flares than modern Earth receives from our Sun in quiescence; all other TRAPPIST-1 planets, being farther away from the star, decrease in flux received as they are successively farther away from the star. Earth receives more UV-B flux now than in the past \citep{Rugheimer2015}, so that each planet receives greater UV-B flux as a proportion of early Earth's UV-B flux than that of modern Earth. However, no planet receives UV-B flare flux anywhere close to that of early Earth. In particular, the habitable-zone planets c, d, and e \citep{Gillon2017} currently receive more than an order of magnitude less quiescent and flare UV-B flux than Earth has ever received in its history. These results are consistent with our results in Sections \ref{subsec:oz_depl} and \ref{subsec:chemmy_bits}, in which we determined that TRAPPIST-1's current bolometric flare energy output does not provide enough UV flux to completely deplete ozone in habitable-zone planet atmospheres, nor enough UV flux to catalyze the prebiotic reactions discussed in \cite{Rimmer2018}.

\section{Conclusions} \label{sec:end}

With three of its seven terrestrial planets in its habitable zone, TRAPPIST-1 is an exciting system to study for clues as to whether its planets might be habitable. Being an extremely cool and red star, TRAPPIST-1 is subject to frequent flaring, and these flares must be considered in studies of its planets due to their effect on planetary atmospheres -- both as harbingers of doom via ozone depletion and atmospheric erosion, or givers of life via light-dependent prebiotic chemistry. Previous work has shown that extreme flare events can destroy ozone columns of planetary atmospheres, but also that cool red stars such as TRAPPIST-1 do not provide enough quiescent UV flux to power UV-dependent prebiotic chemistry necessary for building RNA nucleotides, thus flares could provide the necessary input of energy for abiogenesis to occur. Better constraining TRAPPIST-1's high-energy flare activity is necessary for assessing whether its planets could be habitable.

Stellar activity cycles can lead to over- or underestimates of a star's overall flare activity rate if only short-term observations are available. Our nondetections of flares over two years rule out that TRAPPIST-1 was particularly quiet while \textit{K2} observed it, consistent with previous work demonstrating the unlikelihood of solar-like activity cycles for the star, but it is possible that TRAPPIST-1 was slightly more active than normal during \textit{K2}'s observations, causing its cumulative flare rate to appear higher than it is on average. Further long-term monitoring of the TRAPPIST-1 system is necessary to better refine its overall flare activity rate, and by extension, assess how great a role its flares may play in its planets' possible habitability. In future work, we plan to extend our analysis here to other systems like TRAPPIST-1 using the Evryscope Fast Transient Engine (EFTE) a newly developed multi-purpose automated pipeline, to analyze differential images \citep{Hank2020}.

Combining Evryscope observations of TRAPPIST-1 with legacy \textit{K2} data, we find that TRAPPIST-1's cumulative superflare rate is consistent with the previously estimated value, decreasing from $4.4^{+2.0}_{-0.2}$ to $4.2^{+1.9}_{-0.2}$ superflares per year. The star's high-energy flare rate is not consistent with total ozone depletion for its planets, being far too low to provide frequent enough UV bombardment to do so. However, TRAPPIST-1's flare rate is also not consistent with abiogenesis; its cumulative flare rate does not provide enough UV flux to catalyze UV-dependent reactions necessary for synthesizing RNA nucleotides on Earth. If TRAPPIST-1's habitable-zone planets are currently habitable, they are unlikely to be in danger of ozone depletion from flares. At the same time, however, its habitable-zone planets do not receive enough UV flux from TRAPPIST-1's flares to sustain abiogenesis, making it unlikely in the present that RNA nucleotides could be assembled by the same processes on its planets as they are on Earth.\\

ALG thanks Anna Reine, M. Collins, Stephen Schmidt, and Anna Hughes for insightful discussions that have improved this work. We thank the anonymous referee for their helpful comments.

This research was partially supported by the NSF CAREER grant AST-1555175 and the Research Corporation Scialog grants 23782 and 23822. HC is supported by the NSF GRF grant DGE-1144081. The Evryscope was constructed under NSF/ATI grant AST-1407589.

This paper includes data collected by the \textit{Kepler} mission. Funding for the \textit{Kepler} mission is provided by the NASA Science Mission directorate.

This research made use of Astropy,\footnote[1]{http://www.astropy.org} a community-developed core Python package for Astronomy \citep{astropy:2013, astropy:2018}, and the NumPy, SciPy, and Matplotlib Python modules \citep{numpyscipy, scipy, matplotlib}.

This research was carried out on the traditional lands of the Occaneechi, Haw, and Eno Native American tribes in Orange County, NC.

\textit{Facilities:} CTIO:Evryscope, \textit{Kepler}

\bibliographystyle{aasjournal}
\bibliography{trappist_refs}

\end{document}